%% file: main.tex
\pdfoutput=1
\documentclass[11pt]{article}
\usepackage[preprint]{acl}

\usepackage{times}
\usepackage{latexsym}
\usepackage{amsmath}
\usepackage[T1]{fontenc}
\usepackage{booktabs}
\usepackage{makecell}
\usepackage{url}
\usepackage[utf8]{inputenc}
\usepackage{multirow}
\usepackage{microtype}
\usepackage{inconsolata}
\usepackage{graphicx}
\usepackage{pifont}
\newcommand{\cmark}{\ding{51}}%
\newcommand{\xmark}{\ding{55}}%

\title{OpenSep: Leveraging Large Language Models with Textual Inversion for Open World Audio Separation}

\author{Tanvir Mahmud, Diana Marculescu \\
    Chandra Family Department of Electrical and Computer Engineering \\
  The University of Texas at Austin \\
  \texttt{tanvmirmahmud@utexas.edu},  \texttt{dianam@utexas.edu}}

\begin{document}
\maketitle

\input{sections/abstract}
\input{sections/introduction}
\input{sections/related_works}
\input{sections/methodology}

\input{sections/results}
\input{sections/conclusion}

\bibliography{custom}

\clearpage
\appendix
\input{sections/appendix}

\end{document}

%% file: sections/abstract.tex
\begin{abstract}
Audio separation in real-world scenarios, where mixtures contain a variable number of sources, presents significant challenges due to limitations of existing models, such as over-separation, under-separation, and dependence on predefined training sources. We propose OpenSep, a novel framework that leverages large language models (LLMs) for automated audio separation, eliminating the need for manual intervention and overcoming source limitations. OpenSep uses textual inversion to generate captions from audio mixtures with \textit{off-the-shelf} audio captioning models, effectively parsing the sound sources present. It then employs few-shot LLM prompting to extract detailed audio properties of each parsed source, facilitating separation in unseen mixtures. Additionally, we introduce a multi-level extension of the \textit{mix-and-separate} training framework to enhance modality alignment by separating single source sounds and mixtures simultaneously. Extensive experiments demonstrate OpenSep's superiority in precisely separating new, unseen, and variable sources in challenging mixtures, outperforming SOTA baseline methods. Code is released at \url{https://github.com/tanvir-utexas/OpenSep.git}.
\end{abstract}

%% file: sections/introduction.tex
\section{Introduction}
Audio and music mostly appear in mixtures in real-world containing various audio sources and background noise~\citep{kim2019audiocaps}. Separating clean sources from noisy mixtures have numerous applications in audio processing~\citep{liu2018casa, music1}. Precise separation of clean sound sources require their complete semantic understanding, as well as extensive training on natural mixtures. However, in open world scenarios, audio mixtures may contain a variable number of sources, as well as new, unseen, and possibly noisy sources. Hence, gathering clean sounds from a variety of sources, and modeling exhaustive mixture combinations for training are often impractical, which limits the use of audio separators in practice.

\begin{figure}[t]
    \centering
  \includegraphics[width=0.95\columnwidth]{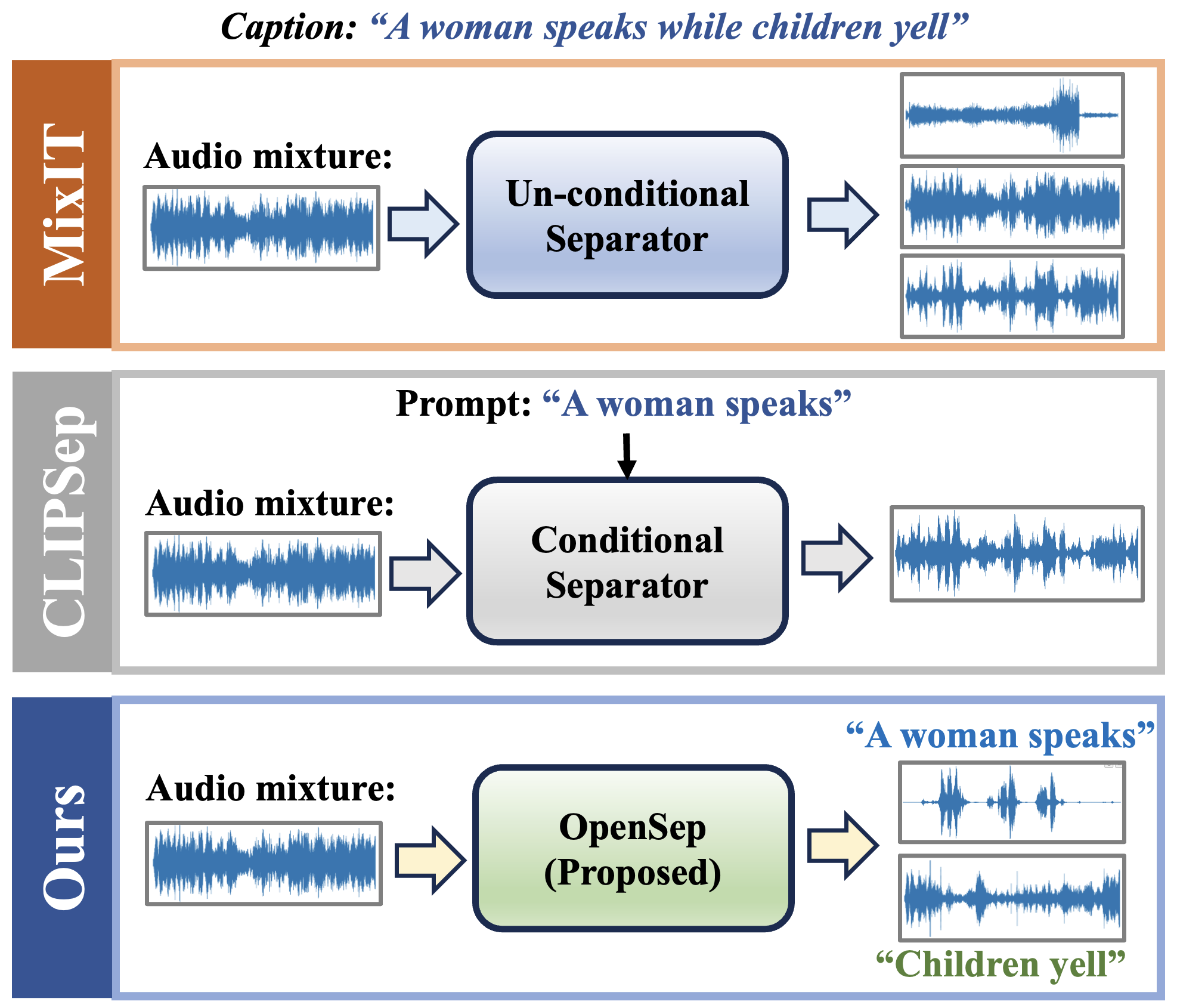}
  \caption{Un-conditional audio separators suffer from both over-separation and under-separation in noisy mixtures, and cannot parse audio entities without additional classifiers. Furthermore, conditional separators rely on manual text prompts for source separation, limiting their use in practice. In contrast, OpenSep fully automates the source parsing and separation flow, even with varying number of unseen and noisy sources in open world.}
  \label{fig:1}
\end{figure}

Prior work on audio separation mostly focused on two approaches: unconditional and conditional source separation. Unconditional separators~\cite{pit, mixit} mostly attempt to disentangle the mixture into a fixed set of output predictions, and later rely on post-processing to select/process the separated sources. However, this approach is largely limited to training sources and mixture combinations, which, in practice, results in over-/under-separation~\citep{mixcycle}. In addition, unconditional separators cannot provide the corresponding class entities of separated sources. 
\citet{weaksup} introduced source classifiers with audio separators, but their method is limited by training sources only.

Conditional source separation simplifies the problem by relying on conditional prompts from other modalities/signals, such as text, vision, and sometimes, clean audio representing the target sound~\citep{sop, mahmud2024weakly}. However, these methods are limited by the availability of target prompts for separation, which are often difficult to gather in practice. Moreover, the conditioning signal usually contains simple class instances as prompts which cannot explicitly describe the target audio source. Hence, these conditional audio separators are mostly limited to seen sources, often under-performing on unseen, noisy source mixtures in open world.

To overcome these limitations, we introduce OpenSep, a novel framework to automatically separate and parse unseen audio from noisy mixtures with a variable number of sources in open world (Fig~\ref{fig:1}). In particular, we propose textual inversion by using an \textit{off-the-shelf} audio captioning model to extract text representations of noisy mixtures. We also leverage the world knowledge of audio sources in large language models (LLMs) to automatically parse, disentangle, and extract enriched representations of audio properties of each source present in the mixture. Finally, we train a text-conditioned audio separator to extract audio sources from noisy mixtures. For enhancing the modality alignment between conditional prompts and separated sources, we propose a multi-order separation objective by extending the baseline \textit{mix-and-separate} framework. Extensive experiments demonstrates significant performance improvements of OpenSep over prior work, \textit{e.g.}, OpenSep achieves $+64\%$ and $+180\%$ SDR improvements on unseen sources in MUSIC and VGGSound datasets, respectively, over baseline state-of-the-art models.

Our contributions are summarized as follows:
\begin{enumerate}
    \item Our work is the first to introduce knowledge parsing from large language models for open-world audio separation.
    \item OpenSep fully automates the source separation and recognition pipeline from noisy, unseen mixtures without manual intervention.
    \item We propose a multi-level extension of \textit{mix-and-separate} training framework to enhance audio and text modality alignment.
    \item Extensive experiments on three benchmark datasets show the superiority of OpenSep over existing state-of-the-art methods.
\end{enumerate}

%% file: sections/related_works.tex
\section{Related Work}

\paragraph{Unconditional Sound Separation}
Prior work on unconditional audio separation in speech and music mostly relies on post-processing methods to pick the target sound~\citep{music1, music2, music3, wang, pit, adl, tasnet}. Later, permutation invariant training (PIT)~\citep{pit, universal}, followed by its variant mixture invariant training (MixIT)~\citep{mixcycle, mixit, mixitv2} relies on permutation alignment on source prediction for performance enhancement. However, these methods suffer from both over- and under-separation, due to training distribution misalignment in open world scenarios. Later, weakly supervised training with classifiers has been explored~\citep{weaksup, audioscope}, however, such methods are limited in their use on a fixed number of training sources. In contrast, OpenSep attempts to separate a variable number of sources in open world, without being limited to certain training sources.

\paragraph{Conditional Sound Separation}
Prior work on conditional sound separation used visual guidance~\citep{cosep, sop, cyclic, scene, matching}, text guidance~\citep{clipsep, lass, kilgour2022text, textnew, anything, mahmud2024weakly}, and clean audio source guidance~\cite{audio1, audio2} for conditioning on noisy mixtures. Most of these methods mostly rely on a \textit{mix-and-separate} framework~\citep{sop}. However, the requirement for users to explicitly specify which sources to separate is often impractical in dynamic or complex audio scenes. Moreover, in general, these methods struggle with unseen, new sources for learning the conditional guidance with specific class prompts. In contrast, OpenSep attempts to fully automate the separation of all sources present in noisy, unseen source mixtures in open world, without using hand crafted prompts.

%% file: sections/methodology.tex
\section{Methodology}
\begin{figure*}[t]
    \centering
  \includegraphics[width=0.95\textwidth]{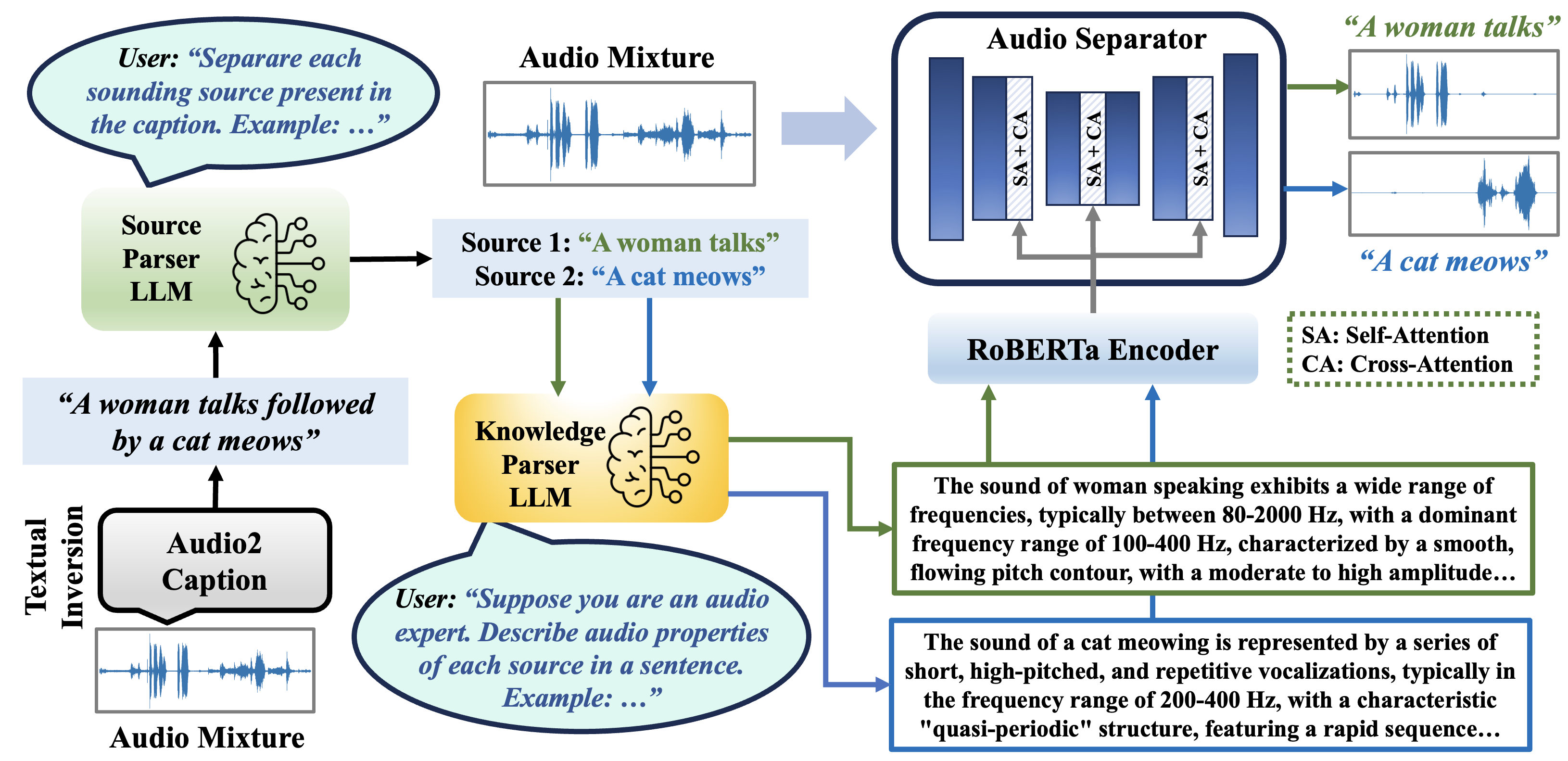}
  \caption{Proposed OpenSep pipeline: We initially apply textual inversion on noisy audio mixtures with an \textit{off-the-self} audio captioning model to extract text descriptions. Afterwards, a pre-trained instruction-tuned LLM is used to parse audio sources from the caption, followed by the extraction of detailed audio properties of each source. Finally, a text-conditional audio-separator is used for separating each audio source from the noisy mixture using the enriched text prompts. Here, the audio separator is trained for leveraging detailed audio properties in textual representation.}
  \label{fig:2}
\end{figure*}

OpenSep addresses two critical challenges of audio separation: handling a variable number of sources without manual intervention and enhancing performance on unseen sources during inference. As illustrated in Fig~\ref{fig:2}, the OpenSep architecture comprises of three key components. First, to overcome the training and test distribution misalignment of the unconditional separator in real-world mixtures, while also eliminating the need for manual prompts in conditional source separators, OpenSep leverages {textual inversion} to convert the audio mixture into a text representation. This step addresses the challenge of detecting a variable number of sources, while eliminating the human intervention to apply text conditions. Following that, we introduce knowledge parsing for each sounding source from large language models. We propose few-shot prompting with instruction-tuned LLM for parsing various sound sources from the caption, followed by the extraction of detailed audio properties of each sound source in text representation. This step is important for performing complex audio mixture separation in open-world scenarios on new, unseen sources. Finally, OpenSep uses a text-conditional audio-separator to extract the target sources based on the LLM-parsed text description. Unlike prior methods, OpenSep relies on detailed LLM-parsed knowledge to gather the context of each sound source for generalization, without being overfitted to training sources, which makes it suitable for real-world applications.

The following sections detail each component of the OpenSep framework to achieve robust and scalable audio separation.

\begin{figure*}[t]
    \centering
  \includegraphics[width=0.88\textwidth]{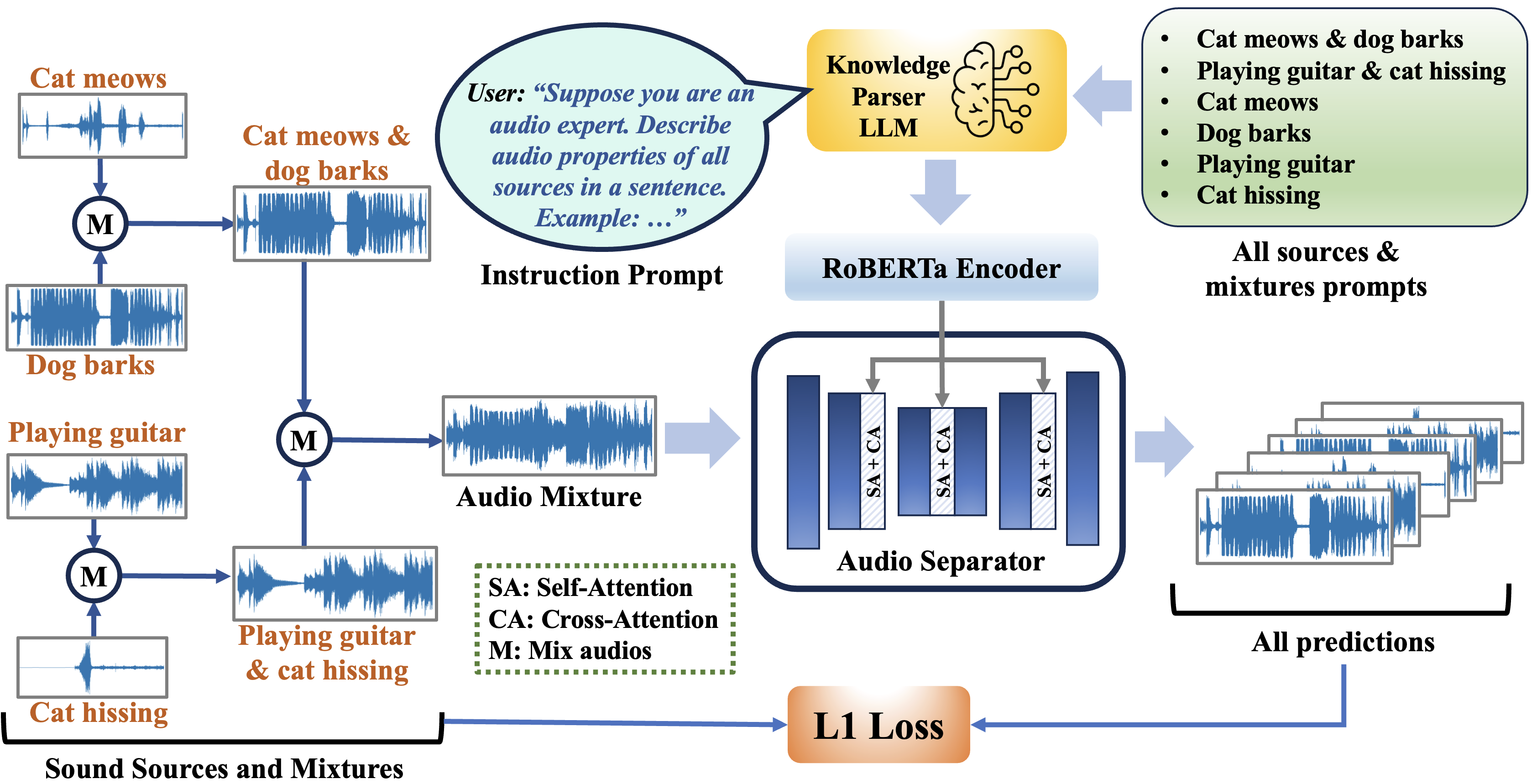}
  \caption{Proposed training pipeline: We extend the baseline \textit{mix-and-separate} framework with multi-order separation objective for enhanced modality-alignment. Initially, we sample four independent single source sounds, and prepare synthetic mixtures of two and four sources. We parse enriched text prompts for mixtures and single-source sounds with a knowledge parser LLM. The audio separator is trained to separate both single-source and lower-order mixtures based on enriched text guidance using an L1 loss objective.}
  \label{fig:3}
\end{figure*}

\subsection{Source Parsing with Textual Inversion}
The audio source separation task can be split into two key phases: (1) detecting the sound sources present in the mixture, and (2) separating each source correctly. Prior work on unconditional separators attempts to perform both tasks simultaneously. This approach is challenging since it attempts to align the data distribution on varying number of sources, often resulting in over/under separation.  Conditional separators achieve superior performance by eliminating the source recognition phase, but they rely heavily on manual prompts, which limits their applicability to automatically parsing the sources present in the mixture. OpenSep solves this complex mixture disentanglement challenge in open world by using textual inversion and leveraging the world knowledge in large language models.

\paragraph{Textual Inversion:} To simplify the source parsing in complex mixtures, we propose converting the audio mixture to a text representation by using an \textit{off-the-shelf} audio captioning model. This model processes the audio input and generates a textual description or caption, which encapsulates the salient features and sources present in the mixture. For example, a caption might describe an audio mixture as "a person speaking with background music and occasional dog barks." This textual representation enables the subsequent use of LLMs to further analyze and extract detailed information about the individual sound sources present in the audio. 

\paragraph{Source parser LLM:} Parsing audio sources requires semantic understanding of each source, which is difficult to train for the dynamic use cases in audio modality, limiting the practical use of audio separators. To parse each source, we propose to use a source parser LLM on the inverted text captions of the raw audio. By leveraging the semantic world knowledge of LLMs and few-shot prompting, OpenSep can precisely extract sound sources from mixtures. We use an instruction-tuned LLM, with the instruction for parsing individual sound sources present in the mixture. For example, with the input caption, \textit{"Children yelling while a dog is barking in the background and a car horn honks afterwards"}, LLM splits each sources as \textit{"Children yelling. Dog barking. Car horn honking."} 

\subsection{Knowledge Parsing for Each Source}
To precisely separate each parsed source from complex mixtures, it is necessary to have more detailed knowledge of audio properties of the target source. Traditional conditional separators only rely on the class representation of each source. Despite their promising performance on seen classes used for training, these models underperform on unseen classes. To overcome this limitation, we propose the use of an instruction-tuned LLM as a knowledge parser to incorporate detailed audio properties of each source prior to the separation. This approach enables denser text alignment via separation of unseen sources in open-world scenarios. To facilitate the knowledge parsing from the LLM, we use specific instruction prompts to utilize the LLM as an audio expert providing the detailed audio properties of each source mentioned. Furthermore, we focus on several key properties of each sounding source for enriched representation, such as frequency range (pitch), amplitude (loudness), timbre (tone quality), usual duration, attack and decay, dynamic envelope characteristics, spectral content detailing the presence of harmonics, overtones, and the overall shape of the sound spectrum. Several hand-crafted high quality prompts are designed for several sources, and these are used for few-shot prompting with instruction guidance to leverage the pre-trained LLMs as an audio expert.

Rather than using a single class name for each sounding instance, we integrate richer details of each sounding source by leveraging knowledge parser LLMs, which in general, enhances audio separation performance by focusing on detailed audio feature guidance. Such denser audio-language alignment plays a pivotal role in  separating unknown and noisy sources in real world.

\subsection{Text-Conditioned Audio Separator}
OpenSep uses a text-conditioned audio separator as a core building block given its superior performance over unconditional separators. The key difference of OpenSep over prior conditional separators is the automatic extraction of enriched audio features via modality inversion. Therefore, a longer context window is used for encoding the text prompts in the audio separator, to align the separator with enhanced text prompts. For the audio separator, we use the U-Net architecture as in prior work. For denser alignment between separated audio features and given text prompts, we use self-attention followed by cross-attention layers after each building block in the U-Net. The model initially converts the audio mixture into a magnitude spectrogram using short-term Fourier transform (STFT), and predicts the mask of the target sound given the text prompt using simple 2D convolutional kernels followed by self-attention and cross-attention. To reconstruct the sound, we use the filtered spectrogram with the phase residuals extracted from the original mixture, following prior work~\citep{clipsep}.

\subsection{Proposed Training Pipeline}
Most prior work relies on a \textit{mix-and-separate} framework for conditional separator, which learns to separate single sources from synthetic audio mixtures given a conditional prompt.
OpenSep primarily focuses on separating the target source by using richer text conditioning rather than a simple class prompt. The overall performance improvement on unseen and noisy sources mostly comes from the deeper audio and textual feature grounding on diverse audio properties. To achieve this, we propose a two-level separation training objective, extending the baseline \textit{mix-and-separate} framework, such as mixture, and single-source separation.

As illustrated in Fig 3, we initially sample four single source audios $(x_1, x_2, x_3, x_4)$ and prepare two synthetic mixtures $(y_1, y_2)$ by mixing each pair of sources, given by $y_1 = \mathtt{Mix}(x_1, x_2)$, and $y_2 = \mathtt{Mix}(x_3, x_4)$. For mixing, we use amplitude re-scaling of each source, followed by simple addition on raw audio waveforms. Afterwards, we generate a higher order mixture $z$ of four sources using $z = \mathtt{Mix}(y_1, y_2)$. By leveraging the instruction tuned LLM on class entities of each source, we extract single source text prompts $(S_1, S_2, S_3, S_4)$, and two-source mixture prompts $(M_1, M_2)$. The model generates a series of predictions $\mathbf{P}$ of separated audios given the text prompt $\mathcal{T}$ and higher-order mixture $z$, where $\mathcal{T} \in \{S_1, S_2, S_3, S_4, M_1, M_2 \}$. Finally, the $L_1$ loss between predicted magnitude spectrogram and source spectrogram is used as training objective.

The proposed multi-level extension of \textit{mix-and-separate} training method facilitates the separation of lower-order audio mixtures as well as single source sounds from higher-order mixtures, following the detailed text feature guidance from the LLM. Hence, this method enhances the modality alignment between separated sounds and enriched LLM-generated text prompts, facilitating the separation on unseen and noisy sources.

\input{tables/seen_results}

\input{tables/unseen_results}

%% file: tables/seen_results.tex
\begin{table*}[]
\centering
\caption{Performance comparison on seen classes. All classes are used for training in each dataset. Baseline results are reproduced under the same setup for a fair comparison. Our method outperforms both conditional and unconditional baselines, without accessing the source prompts during evaluation.}
\label{tab:seen}
\scalebox{0.85}{
\begin{tabular}{ccccc}
\toprule
\multirow{2}{*}{Methods} & \multicolumn{2}{c}{MUSIC} & \multicolumn{2}{c}{VGGSound} \\
\cmidrule(lr){2-3} \cmidrule(lr){4-5}
                         & SDR $\uparrow$         & SIR $\uparrow$        & SDR $\uparrow$           & SIR $\uparrow$       \\
\midrule
PIT~\citep{pit}                      & 7.98 \small $\pm$ 0.27 & 10.81 \small $\pm$ 0.29 & 2.01 \small $\pm$ 0.32   & 4.13 \small $\pm$ 0.27  \\
MixIT~\citep{mixit}                    & 5.46 \small $\pm$ 0.42 & 7.39 \small $\pm$ 0.22 & 1.46 \small $\pm$ 0.19   & 3.56 \small $\pm$ 0.25  \\
MixPIT~\citep{mixcycle}                      & 8.07 \small $\pm$ 0.25 & 11.01 \small $\pm$ 0.28 & 1.87 \small $\pm$ 0.26   & 3.95 \small $\pm$ 0.31  \\
MixIT + PIT              & 8.13 \small $\pm$ 0.28 & 11.17 \small $\pm$ 0.23 & 2.14 \small $\pm$ 0.24   & 4.98 \small $\pm$ 0.41  \\
\midrule
CLIPSep~\citep{clipsep}                  & 8.33 \small $\pm$ 0.29 & 11.65 \small $\pm$ 0.26 & 2.24 \small $\pm$ 0.21   & 5.41 \small $\pm$ 0.23  \\
WeakSup~\citep{weaksup}                  & 6.49 \small $\pm$ 0.35 & 8.87 \small $\pm$ 0.31 & 1.73 \small $\pm$ 0.34   & 4.67 \small $\pm$ 0.29  \\
LASSNet~\citep{lass}                  & 8.35 \small $\pm$ 0.32 & 11.83 \small $\pm$ 0.29 & 2.32 \small $\pm$ 0.29   & 5.95 \small $\pm$ 0.34  \\
AudioSep~\citep{liu2023separate}                 & 8.64 \small $\pm$ 0.31 & 12.18 \small $\pm$ 0.27 & 2.45 \small $\pm$ 0.23   & 6.14 \small $\pm$ 0.31  \\
\midrule
\textbf{OpenSep (Ours)}                 & \textbf{9.56} \small $\pm$ 0.28 & \textbf{13.42} \small $\pm$ 0.26 & \textbf{3.71} \small $\pm$ 0.18   & \textbf{8.31} \small $\pm$ 0.19 \\
\bottomrule
\end{tabular}}
\end{table*}

%% file: tables/unseen_results.tex
\begin{table*}[t]
\centering
\caption{Performance comparison on unseen classes. All  models are trained with 50\% class samples, and the evaluation is carried out on remaining 50\% class mixtures. OpenSep demonstrates significantly better performance on new, unseen classes compared to existing baselines.}
\label{tab:unseen}
\scalebox{0.85}
{\begin{tabular}{ccccc}
\toprule
\multirow{2}{*}{Methods} & \multicolumn{2}{c}{MUSIC} & \multicolumn{2}{c}{VGGSound} \\
\cmidrule(lr){2-3} \cmidrule(lr){4-5}
                         & SDR $\uparrow$        & SIR $\uparrow$          & SDR $\uparrow$            & SIR $\uparrow$           \\
\midrule
PIT~\citep{pit}                      & 3.56 \small $\pm$ 0.29 & 4.97 \small $\pm$ 0.33 & 0.45 \small $\pm$ 0.29   & 1.54 \small $\pm$ 0.33  \\
MixIT~\citep{mixit}                    & 2.28 \small $\pm$ 0.35 & 3.45 \small $\pm$ 0.29 & 0.16 \small $\pm$ 0.33   & 0.97 \small $\pm$ 0.21  \\
MixPIT~\citep{mixcycle}                      & 2.87 \small $\pm$ 0.26 & 4.11 \small $\pm$ 0.35 & 0.28 \small $\pm$ 0.31   & 1.13 \small $\pm$ 0.35  \\
MixIT + PIT              & 3.98 \small $\pm$ 0.26 & 5.25 \small $\pm$ 0.24 & 0.66 \small $\pm$ 0.24   & 2.15 \small $\pm$ 0.34  \\
\midrule
CLIPSep~\citep{clipsep}                  & 4.97 \small $\pm$ 0.27 & 7.13 \small $\pm$ 0.24 & 1.08 \small $\pm$ 0.27   & 4.12 \small $\pm$ 0.29  \\
WeakSup~\citep{weaksup}                  & -3.56 \small $\pm$ 0.47 & -4.36 \small $\pm$ 0.37 & -4.54 \small $\pm$ 0.52   & -5.86 \small $\pm$ 0.89  \\
LASSNet~\citep{lass}                  & 5.01 \small $\pm$ 0.23 & 7.38 \small $\pm$ 0.29 & 0.95 \small $\pm$ 0.26   & 3.89 \small $\pm$ 0.35  \\
AudioSep~\citep{liu2023separate}                 & 5.14 \small $\pm$ 0.24 & 7.56 \small $\pm$ 0.29 & 1.12 \small $\pm$ 0.42   & 4.45 \small $\pm$ 0.27  \\
\midrule
\textbf{OpenSep (Ours)}                  & \textbf{8.45} \small $\pm$ 0.32 & \textbf{11.72} \small $\pm$ 0.35 & \textbf{3.14} \small $\pm$ 0.31   & \textbf{7.23} \small $\pm$ 0.39 \\
\bottomrule
\end{tabular}}
\end{table*}

%% file: sections/results.tex
\section{Results}

\subsection{Evaluation Setup}
\paragraph{Dataset:}
For the experiments on synthetic mixtures, we primarily use MUSIC~\citep{sop} and VGGSound~\citep{chen2020vggsound} datasets. MUSIC dataset contains 21 musical instruments, separately played and recorded for $1\sim 5$ minutes duration. VGGSound contains $162,433$ audio samples, mostly containing noisy single source sounds of $10$s duration. For analyzing the performance on natural mixtures, we use AudioCaps~\citep{kim2019audiocaps} dataset containing $44,309$ audio mixtures having around $1\sim6$ sources with audio captions. 

\paragraph{Implementation Details:}
We use LLaMA-3-8b~\citep{touvron2023llama} language model for source and knowledge parsing. We use RoBERTa-Base~\citep{liu2019roberta} text encoder with a context window of $512$ for encoding parsed LLM knowledge. For audio captioning, we use the CLAP-based captioning model~\citep{elizalde2023clap}, which combines an audio encoder with a GPT-2 text decoder. We use a U-Net based audio separator with self and cross-attention conditioning.

\paragraph{Training:} All models are trained for $80$ epochs with initial learning rate of $0.001$. The learning rate is decreased by a factor of $0.1$  every $20$ epochs. An Adam optimizer is used with $\beta_1 = 0.9$, $\beta_2 = 0.999$ and $\epsilon = 10^{-8}$. The training was carried out with $8$ RTX-A6000 GPUs with $48$GB memory. For training, we randomly sample different sources, then, mix and separate using our training method. Also, the class label of each source is used for knowledge parsing during training.

\paragraph{Evaluation:} We use signal-to-distortion ratio (SDR) and signal-to-interference ration (SIR)~\citep{metric} for evaluating different models. In general, SDR estimates overall separation quality of the target sound, which is widely used in prior work~\citep{clipsep, mahmud2024weakly}. SIR estimates the amount of interference in separation from other sources present in the mixture.

\begin{figure*}[t]
    \centering
  \includegraphics[width=\textwidth]{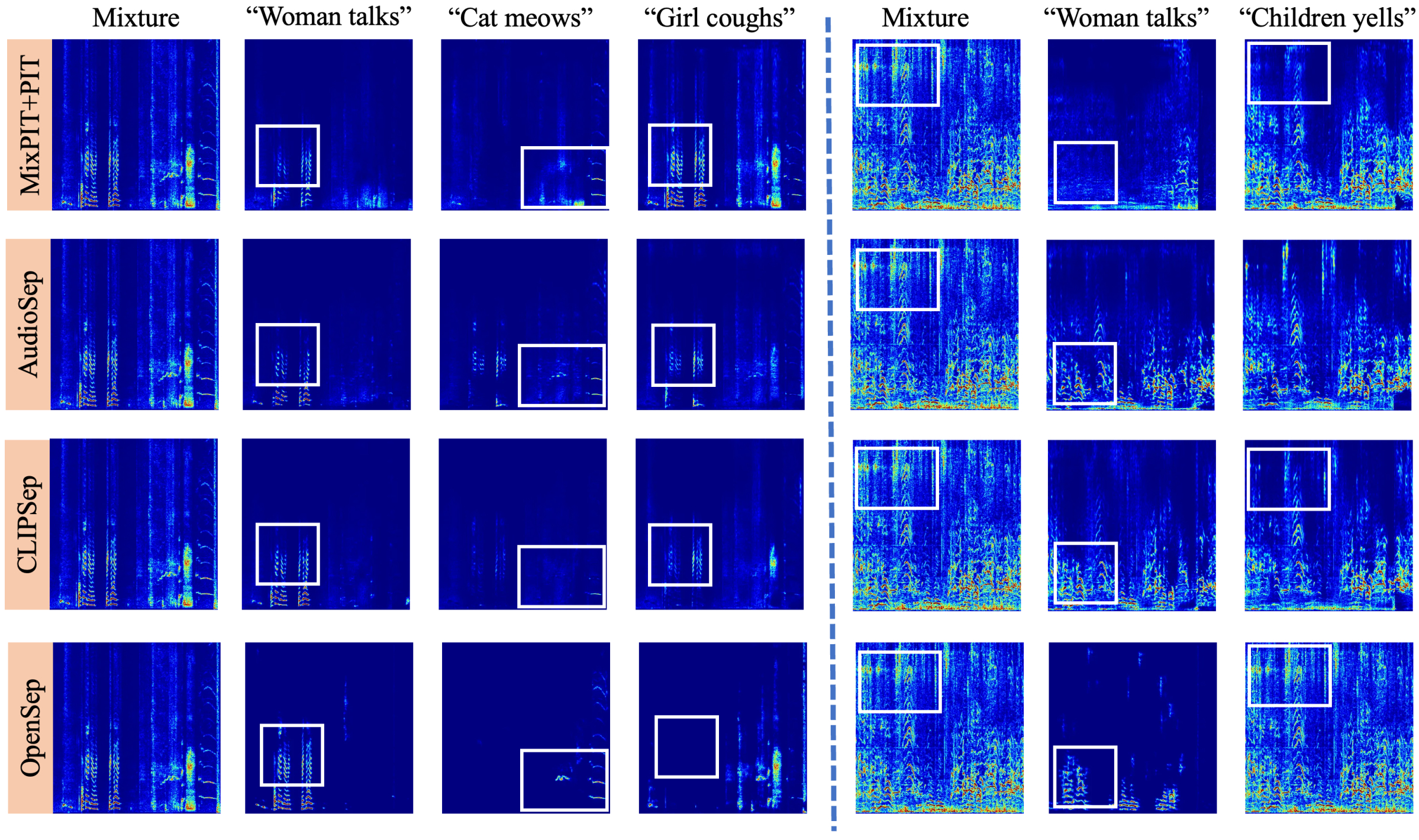}
  \caption{Qualitative results on natural mixtures from AudioCaps. (Left) All baselines show large spectral overlap of \textit{woman talking} sound in other two source predictions. OpenSep precisely disentangles all three sources minimizing the spectral overlap across sources, while preserving spectral details. (Right) For the dominant noisy sound of \textit{children yelling}, all baselines can hardly separate the \textit{woman talking} sound. OpenSep significantly reduces noise in \textit{woman talking}, while preserving spectral details of noisy \textit{children yelling} sound.}
  \label{fig:qual1}
\end{figure*}

\subsection{Main Comparisons}
\paragraph{Baseline Models:}
We used single-source permutation invariant training, PIT~\citep{pit} and multi-source MixIT~\cite{mixit, mixitv2} for unconditional baselines. Note that we use synthetic mixtures for MixIT training. We also combine MixIT and PIT to train on both mixtures and single source sounds.
For conditional baselines, we use LASSNet~\cite{lass}, AudioSep~\cite{liu2023separate}, and CLIPSep~\citep{clipsep} models with text conditions, which are built using the \textit{mix-and-separate}~\citep{sop} framework.
For a fair comparison, all baseline results are reproduced with similar architecture and data split. For the conditional models, we explicitly provide the class labels as conditions. For the OpenSep and unconditional baselines, no external text conditions are used, and the best match between prediction and ground truth source is used for evaluation.

\paragraph{Comparison on Seen Classes:}
In Table~\ref{tab:seen}, we provide a comparison on seen classes. For this analysis, we generate a synthetic test set of two-source mixtures  with 10\% samples from each class. All training was done separately on each dataset on all  classes. We note that unconditional models suffer from over- and under-separation causing lower performance. In general, conditional models achieve superior performance over unconditional models by leveraging the user-defined class prompts. In VGGSound dataset, OpenSep improves by $+99\%$ SDR over MixPIT~\citep{mixcycle} and $+52\%$ SDR over AudioSp models, without accessing text prompts. Moreover, we particularly observe large increase in SIR demonstrating significantly less interference in OpenSep, due to the better disentanglement of corresponding sources. 

 \input{tables/user_preference}
\input{tables/ablation}
\input{tables/few_shot}
\input{tables/llms}
\input{tables/caption_ablate}

\paragraph{Generalization to Unseen Classes:}
For this analysis in Table~\ref{tab:unseen}, only $50\%$ classes from each datasets are used for training, while the test set is formed with the remaining $50\%$ classes in each set. Unconditional separators largely struggle in separating unseen sounds. Conditional separators comparatively achieve superior performance, however, we observe significant performance drop from seen classes. For example, CLIPSep experiences $40\%$ and $54\%$ SDR drops in MUSIC and VGGSound, respectively, while OpenSep achieves significantly higher performance over all baselines, with minimal performance SDR drops of $8\%$ and $15\%$.

\paragraph{Comparison on higher order natural mixtures:}
We use natural mixtures from AudioCaps dataset containing $2~\sim 3$ sources. We train all models on VGGSound dataset and evaluate their performance on AudioCaps to mimic real-world use. For conditional models, we manually extract the condition prompts of all sources using the ground truth caption. Since we don't have access to ground truth sources for evaluation on natural mixtures, we perform human evaluations on the separated sounds given in Tab.~\ref{tab:user_study_details}. OpenSep wins over conditional models in separating all sources, even without using any conditions (71\% \textit{vs.} 3.8\% excluding draws). 
When compared with unconditional models such as MixIT, OpenSep wins over $95\%$ (\textit{vs.} 1.2\% excluding draws) cases denoting its robustness. We also provide some qualitative results in Fig.~\ref{fig:qual1} (see more in appendix). Of note, OpenSep precisely detects, disentangles, and preserves the details in source separation from complex mixtures.

\subsection{Ablation Study}
We use VGGSound dataset in both seen and unseen cases for the ablation study.

\paragraph{Effect of knowledge parsing:}
We study the performance of OpenSep with and without knowledge parsing (Tab.~\ref{t4}). We observe that, by injecting details of audio properties through knowledge parsing, we achieve  SDR improvements of $+33\%$ and  $+98\%$, on seen and unseen classes, respectively, highlighting the effectiveness of our approach.

\paragraph{Effect of proposed training method:}
We ablate the proposed multi-level extension of \textit{mix-and-separate} framework (Tab.~\ref{t4}). We observe large improvements of $12\%$ and $22\%$ on seen and unseen classes, respectively, by using the proposed extension of \textit{mix-and-separate} training. This shows its effectiveness in enhancing audio-language alignment for better performance.

\paragraph{Effect of few shot prompting:}
To guide LLM model parsing the salient audio properties, some high quality manually designed prompts are used. We ablate the effect of few shot prompting (see Tab.~\ref{t4},~\ref{t5}). We observe notable performance gain with few shot prompting by guiding the LLM to extract required details.

\paragraph{Ablation of LLMs:}
We ablate the context from various open-source LLMs, such as LLaMA-2-7b, LLaMA-3-8b~\citep{touvron2023llama}, Mistral-7b~\citep{jiang2023mistral}, Phi-3-medium~\citep{abdin2024phi}, and Gemma-7b~\citep{team2024gemma}, while keeping the context length of 512 (See Tab.~\ref{t6}). Though we found competitive results across LLMs, LLaMA-3-8b generates best results.

\paragraph{Accuracy of LLM Parser and captioning:}
To estimate the accuracy of the LLM parser with several captioning models, such as ms-CLAP~\citep{elizalde2023clap}, fine-tuned Whisper-Large~\citep{kadlvcik2023whisper}, and ACT-Large~\cite{Mei2021act}. We use synthetic mixtures from VGGSound dataset for this analysis. Since each sample contains significant background noise which is not present in class labels, we only consider the top predictions with highest similarity with source labels in mixtures. We use a CLAP-text encoder to estimate the similarity between parsed source texts and ground truth labels. 
We observe considerably higher accuracy with ms-CLAP captioning model compared to other models during source parsing.

%% file: tables/user_preference.tex
\begin{table}[t]
\centering
\caption{\textbf{User preference study on natural mixture separation}. OpenSep demonstrates superior separation quality without accessing conditional prompts.  \label{tab:user_study_details}}
\scalebox{0.77}{
\begin{tabular}{@{}lccc@{}}
\toprule
Model & \makecell{OpenSep wins \\ (mean $\pm$ std)} & \makecell{Draws \\  (mean $\pm$ std)} & \makecell{OpenSep loses \\ (mean $\pm$ std)} \\ \midrule
vs. PIT & $ 81.2 $ \small $ \pm  4.3$ & $17.3 $ \small $ \pm  6.5$ & $1.5 $ \small $ \pm  3.8$ \\
vs. MixIT           & $95.6 $ \small $ \pm  3.2$ & $3.2 $ \small $ \pm  4.9$ & $1.2 $ \small $ \pm  4.4$ \\
vs. CLIPSep        & $75.8 $ \small $ \pm  8.5$ & $20.4 $ \small $ \pm  7.8$  & $3.8 $ \small $ \pm  5.6$ \\
vs. AudioSep         & $65.8 $ \small $ \pm  7.3$  & $30.9 $ \small $ \pm  8.5$  & $3.3 $ \small $ \pm  6.6$ \\
vs. LASSNet       & $69.7 $ \small $ \pm  8.1$  & $27.4 $ \small $ \pm  6.8$  & $2.9 $ \small $ \pm  4.8$ \\ \bottomrule
\end{tabular}}
\end{table}

%% file: tables/ablation.tex
\begin{table*}[t]
\centering
\caption{Ablation study of proposed building blocks. The combination of LLM-knowledge parsing with few-shot prompting, and multi-level extension of mix-and-separate contribute to the best performance in OpenSep. }
\label{t4}
\scalebox{0.82}{
\begin{tabular}{ccccccc}
\toprule
\multirow{2}{*}{\textbf{\begin{tabular}[c]{@{}c@{}}Knowledge\\ Parsing\end{tabular}}} & \multirow{2}{*}{\textbf{\begin{tabular}[c]{@{}c@{}}Few-shot\\ Prompting\end{tabular}}} & \multirow{2}{*}{\textbf{\begin{tabular}[c]{@{}c@{}}Multi-level\\ Training\end{tabular}}} &
\multicolumn{2}{c}{\textbf{Seen Classes}} & \multicolumn{2}{c}{\textbf{Unseen Classes}} \\
\cmidrule(lr){4-5} \cmidrule(lr){6-7}
                                                                                      &                                                                                        &                                                                                          & \textbf{SDR} $\uparrow$        & \textbf{SIR} $\uparrow$        & \textbf{SDR} $\uparrow$         & \textbf{SIR} $\uparrow$         \\
\midrule
\xmark                                                                                     & \xmark                                                                                      & \xmark                                                                                        & 2.19 \small $\pm$ 0.27         & 5.25 \small $\pm$ 0.29         & 1.01 \small $\pm$ 0.31          & 4.13 \small $\pm$ 0.24          \\
\xmark                                                                                     & \xmark                                                                                      & \cmark                                                                                        & 2.78 \small $\pm$ 0.23         & 6.45 \small $\pm$ 0.23         & 1.95 \small $\pm$ 0.34          & 5.59 \small $\pm$ 0.32          \\
\cmark                                                                                     & \xmark                                                                                      & \xmark                                                                                        & 2.92 \small $\pm$ 0.31         & 6.84 \small $\pm$ 0.27         & 2.06 \small $\pm$ 0.29          & 6.01 \small $\pm$ 0.29          \\
\cmark                                                                                     & \cmark                                                                                      & \xmark                                                                                        & 3.26 \small $\pm$ 0.29         & 7.38 \small $\pm$ 0.22         & 2.56 \small $\pm$ 0.27          & 6.69 \small $\pm$ 0.33          \\
\cmark                                                                                     & \cmark                                                                                      & \cmark                                                                                        & \textbf{3.71} \small $\pm$ 0.22         & \textbf{8.31} \small $\pm$ 0.19         & \textbf{3.14} \small $\pm$ 0.31          & \textbf{7.23} \small $\pm$ 0.39         \\
\bottomrule
\end{tabular}}
\end{table*}

%% file: tables/few_shot.tex
\begin{table}[]
\centering
\caption{Ablation of few-shot prompts in instruction-guided knowledge parsing from LLM.}
\label{t5}
\scalebox{0.75}
{\begin{tabular}{ccccc}
\toprule
\multirow{2}{*}{\textbf{\begin{tabular}[c]{@{}c@{}}Sample\\ Shots\end{tabular}}} & \multicolumn{2}{c}{\textbf{Seen Classes}} & \multicolumn{2}{c}{\textbf{Unseen Classes}} \\
\cmidrule(lr){2-3} \cmidrule(lr){4-5}
                                                                                   & \textbf{SDR} $\uparrow$        & \textbf{SIR} $\uparrow$        & \textbf{SDR} $\uparrow$         & \textbf{SIR} $\uparrow$         \\
\midrule
1                                                                                 & 3.31 \small $\pm$ 0.29                & 7.63 \small $\pm$ 0.27              & 2.67 \small $\pm$ 0.32                & 6.52 \small $\pm$ 0.26                 \\
2                                                                                 & 3.52 \small $\pm$ 0.31                & 7.94 \small $\pm$ 0.31              & 2.93 \small $\pm$ 0.34                & 6.85 \small $\pm$ 0.29                 \\
3                                                                                & 3.63 \small $\pm$ 0.26                & 8.15 \small $\pm$ 0.28              & 3.06 \small $\pm$ 0.33                & 7.11 \small $\pm$ 0.31                 \\
\textbf{5}                                                                                & \textbf{3.71} \small $\pm$ 0.22                & \textbf{8.31} \small $\pm$ 0.25              & \textbf{3.14} \small $\pm$ 0.31                & \textbf{7.23} \small $\pm$ 0.39                \\ \bottomrule
\end{tabular}}
\end{table}

%% file: tables/llms.tex
\begin{table}[]
\centering
\caption{Ablation of different large language models in knowledge parsing for audio sources.}
\label{t6}
\scalebox{0.7}
{\begin{tabular}{lcccc}
\toprule
\multirow{2}{*}{\textbf{\begin{tabular}[c]{@{}c@{}}LLM\\ Arch.\end{tabular}}} & \multicolumn{2}{c}{\textbf{Seen Classes}} & \multicolumn{2}{c}{\textbf{Unseen Classes}} \\
\cmidrule(lr){2-3} \cmidrule(lr){4-5}
                                                                                   & \textbf{SDR} $\uparrow$        & \textbf{SIR} $\uparrow$       & \textbf{SDR} $\uparrow$        & \textbf{SIR} $\uparrow$        \\
\midrule
LLaMA-3-8b                                                                                 & \textbf{3.71} \small $\pm$ 0.22                & \textbf{8.31} \small $\pm$ 0.19              & \textbf{3.14} \small $\pm$ 0.31                & \textbf{7.23} \small $\pm$ 0.39                 \\
LLaMA-2-7b                                                                                 & 3.58 \small $\pm$ 0.24                & 8.12 \small $\pm$ 0.24              & 3.06 \small $\pm$ 0.33                & 7.11 \small $\pm$ 0.25                 \\
Mistral-7b                                                                                 & 3.63 \small $\pm$ 0.23                & 8.17 \small $\pm$ 0.21              & 3.09 \small $\pm$ 0.25                & 7.15 \small $\pm$ 0.29                 \\
Phi-3-medium                                                                                & 3.55 \small $\pm$ 0.29                & 8.01 \small $\pm$ 0.24              & 3.05 \small $\pm$ 0.29                & 7.12 \small $\pm$ 0.28                 \\
Gemma-7b                                                                                & 3.45 \small $\pm$ 0.31                & 7.83 \small $\pm$ 0.29              & 2.85 \small $\pm$ 0.29                & 6.93 \small $\pm$ 0.24                \\ \bottomrule
\end{tabular}}
\end{table}

%% file: tables/caption_ablate.tex
\begin{table}[t]
\centering
\caption{Analyzing source parsing accuracy with different audio captioning models in complex mixtures.}
\label{t7}
\small
\begin{tabular}{cccc}
\toprule
\textbf{Caption Model} & \textbf{2-Source} & \textbf{3-Source} & \textbf{4-Source} \\
\midrule
ms-CLAP                   & 96.89             & 91.57             & 87.93             \\
Whisper-Large                & 93.47             & 88.09             & 84.68             \\
ACT-Large       & 90.58             & 83.67             & 78.43            \\
\bottomrule
\end{tabular}
\end{table}

%% file: sections/conclusion.tex
\section{Conclusion}
In this paper, we introduced OpenSep, a novel framework for audio source separation in open-world scenarios. In particular, OpenSep leverages an \textit{off-the-shelf} audio captioning model and the world knowledge embedded in large language models (LLMs) to automatically parse and disentangle audio sources from noisy mixtures. By employing a text-conditioned audio separator and a multi-level mixture separation training objective, our method effectively enhances the alignment between conditional prompts and separated sources. Extensive experiments on three benchmark datasets demonstrate the superior performance of OpenSep, which achieves significant SDR improvements over state-of-the-art methods. 
Our work paves the way for practical, automated audio separation, addressing key limitations of existing methods and enabling future research in open-world audio processing.

\section{Limitations}
OpenSep performance is limited by the precise detection of sound sources in noisy mixtures, which mostly stems from the challenge in having a suitable audio captioning model. Nevertheless, OpenSep framework can potentially integrate any superior audio captioning approach to scale-up on real world cases. Finally, given its use of multiple building blocks, OpenSep is computationally expensive in general. However, by further optimizing the architecture, such as using mobile LLMs (\textit{e.g.}, Phi-3-mini, Gemma-2b), OpenSep computational cost can be largely reduced, which we leave for future study.

\section{Ethics Statement}
We only use publily available dataset for this study.

\section*{Acknowledgement}
This work was supported in part by ONR Minerva program, iMAGiNE - the Intelligent Machine Engineering Consortium at UT Austin, and a UT Cockrell School of Engineering Doctoral Fellowship.

%% file: sections/appendix.tex
\newpage

\section{Appendix}

\subsection{Implementation Details}
We use audio segments of $10$s duration with a sampling rate of $16000$ Hz for all experiments. Each audio sample is processed with short-term Fourier transform (STFT) using the frame window of $1022$, and the hop length of $256$. Following prior work~\citep{clipsep, sop}, the supervision is provided on the mask prediction for each source, instead of final reconstruction. We use the similar conditional U-Net based encoder decoder architecture following prior work~\citep{mahmud2024weakly}. The U-Net model contains a total of seven successive encoding and decoding stages with convolutional kernels having $43.2$M parameters. We apply self-attention followed by cross-attention in the skip connection of four bottom feature levels. A multi-head attention~\citep{attention} is used with $8$ heads for each attention operation.
We use LLaMA-3-8b~\citep{touvron2023llama} language model for parsing both sound sources and their corresponding details of audio properties, with $5$-shot prompting for both parsing phases. These examples are manually curated and refined for guiding LLMs in diverse 
scenarios. We use RoBERTa-base language encoder to encode the extracted knowledge from LLM. A single sentence knowledge of audio properties is extracted for each parsed source targeting the maximum context length of $512$. For the evaluation, we use \textit{torch-mir-eval}~\citep{torch_mir_eval} package for estimating both SDR and SIR in source separation from mixtures.

\subsection{Sample of Source Parsing with LLM}
We present several samples of textual inversion with audio captioning and source parsing from real-world audio mixtures collected from AudioCaps~\citep{kim2019audiocaps} in Table~\ref{tab:8}. We use ms-CLAP~\citep{elizalde2023clap} model for audio captioning, followed by instruction-tuned LLaMA-3-8b model for source parsing. In most cases, the generated captions contain all sound sources presented in the audio mixtures. In general, the generated captions simplify the source descriptions compared to the ground truth captions. Nevertheless, by leveraging the detailed knowledge parsing from the LLM, we enrich details of each source. In a few cases of higher order mixtures, we observe the captioning model cannot detect muffled, short duration sounds. Nonetheless, we observe accurate source parsing with the LLM from the generated captions. Though source parsing is a simple objective, however, it can be complicated in some scenarios with complex captions. For example, the "\textit{cat meowing}" sound is detected twice with the generated captions for repeated sounds, though it represents a single source. However, LLM based source parsing effectively solves such challenges.

\subsection{Sample of Knowledge Parsing with LLM}
We provide samples of knowledge parsing from LLM in Table~\ref{tab:9} for various sound sources. We use instruction-tuned LLaMA-3-8b for such parsing. Several key properties of audios, such as frequency range, amplitude, dynamic envelope characteristics, usual duration, and spectral contents are focused by guiding the LLM with manually curated $5$-shot prompts. Such details of audio properties largely help the audio separator model to ground diverse audio features with text description for enhanced separation, particularly in noisy, unseen mixtures.

\subsection{Additional Qualitative Results}
\begin{figure*}[t]
    \centering
  \includegraphics[width=\textwidth]{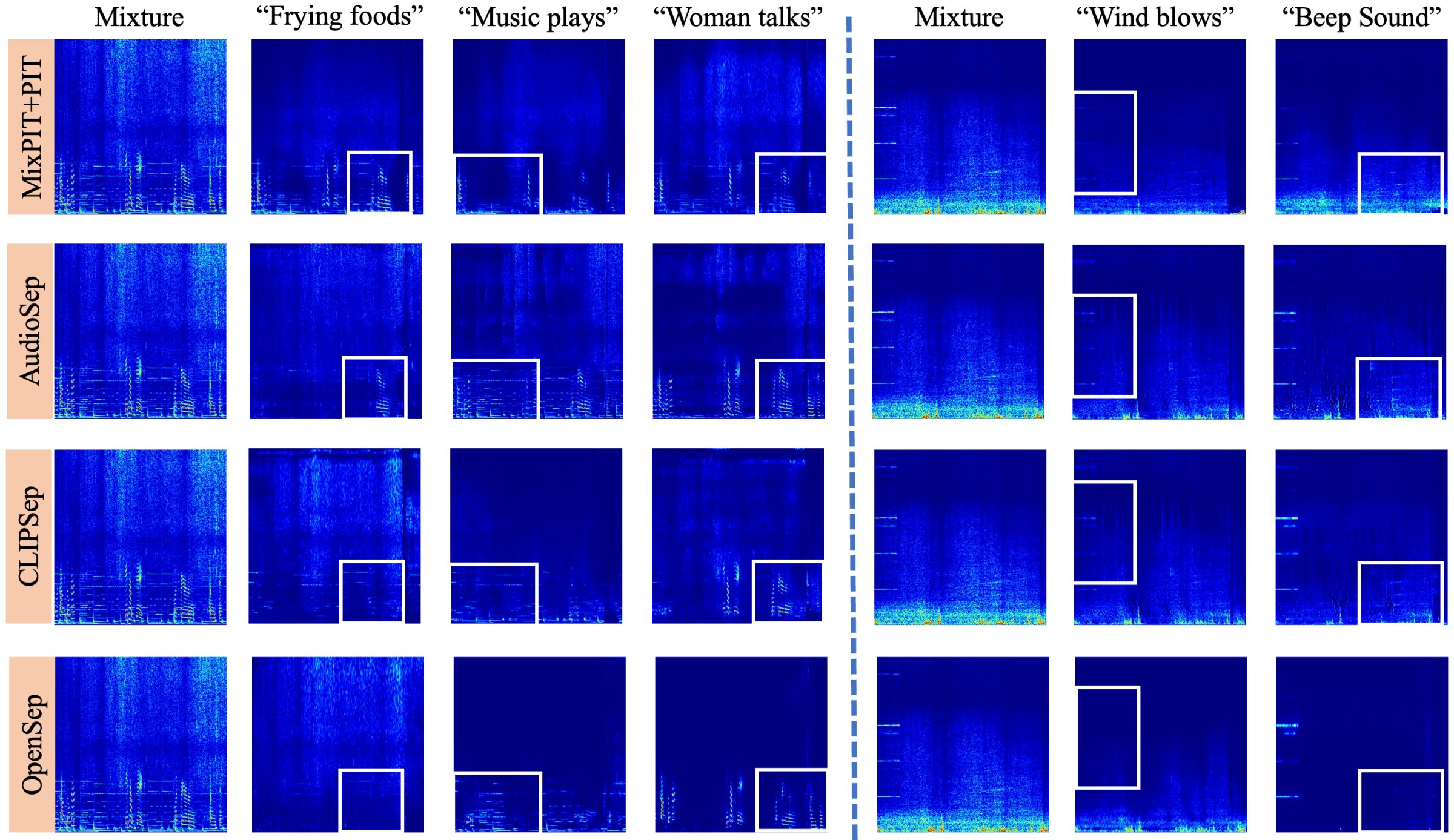}
  \caption{Qualitative results on natural mixtures from AudioCaps. (Left) We can observe the dominant "\textit{woman talks}" spectral content in "\textit{frying foods}" for most baselines. However, in CLIPSep, such overlap is largely reduced, but horizontal spectral contents from "\textit{music plays}" is visible. In contrast, OpenSep largely reduces such spectral overlap in all three components while preserving all details. (Right) In this mixture, the "\textit{beep sound}" is only present at the beginning, with large noisy sound of "\textit{wind blows}" over the spectrogram. Most baseline methods contain noisy spectral contents in the "\textit{beep sound}", while losing spectral contents in the "\textit{wind blows}" prediction. In contrast, OpenSep disentangles this noisy mixture with significant reduction of spectral overlaps.}
  \label{fig:5}
  \vspace{10mm}
\end{figure*}

\begin{figure*}[t]
    \centering
  \includegraphics[width=\textwidth]{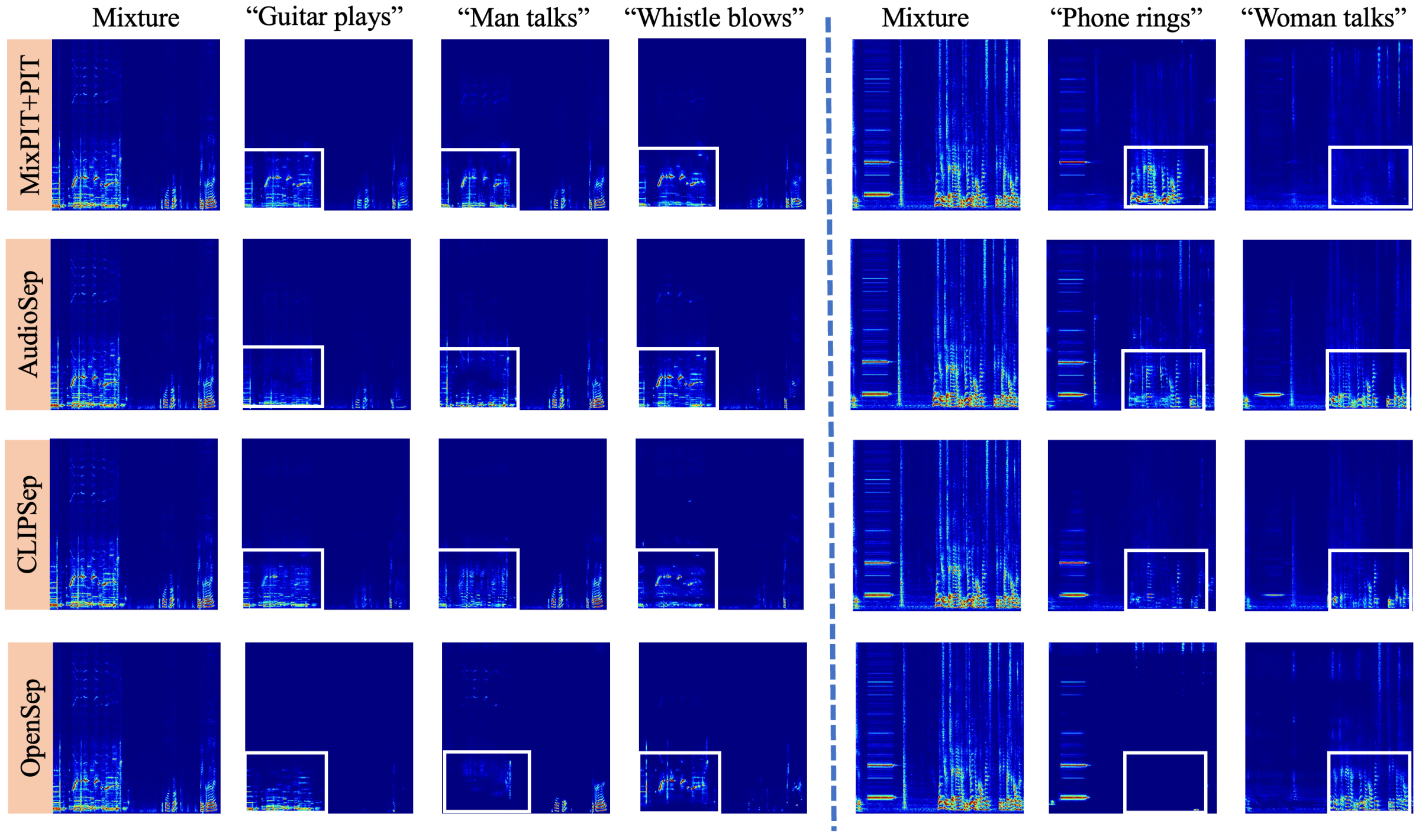}
  \caption{Qualitative results on natural mixtures from AudioCaps. (Left) In the initial phase, the "\textit{whistle blows}" and "\textit{guitar plays}" sounds are present, while the "\textit{man talks}" sound appears at the end. In all baselines, we can see significant spectral overlaps of the "\textit{whistle blows}" and "\textit{guitar plays}" sounds. In contrast, OpenSep precisely separates both of these challenging components, while also reducing background contents in the "\textit{man talks}" prediction. (Right) The "\textit{phone rings}" sound mostly appears at the beginning followed by the "\textit{woman talks}" sound at the end. Compared to all baselines, OpenSep more sharply disentangles both of these sources from this challenging mixture highlighting its effectiveness in practice.}
  \label{fig:6}
  \vspace{10mm}
\end{figure*}

We provide additional qualitative comparisons for natural audio mixture separation in Fig.~\ref{fig:5} and Fig.~\ref{fig:6}. In general, OpenSep largely reduces the spectral overlap across multiple sources while preserving details of each source in separation from challenging mixtures, compared to state-of-the-art baseline methods, without accessing manual source prompts.
We also provide audio samples from different models in the supplementary for better understanding of the separation performance.

\subsection{Details of User Study}
\begin{figure*}[t]
    \centering
  \includegraphics[width=\textwidth]{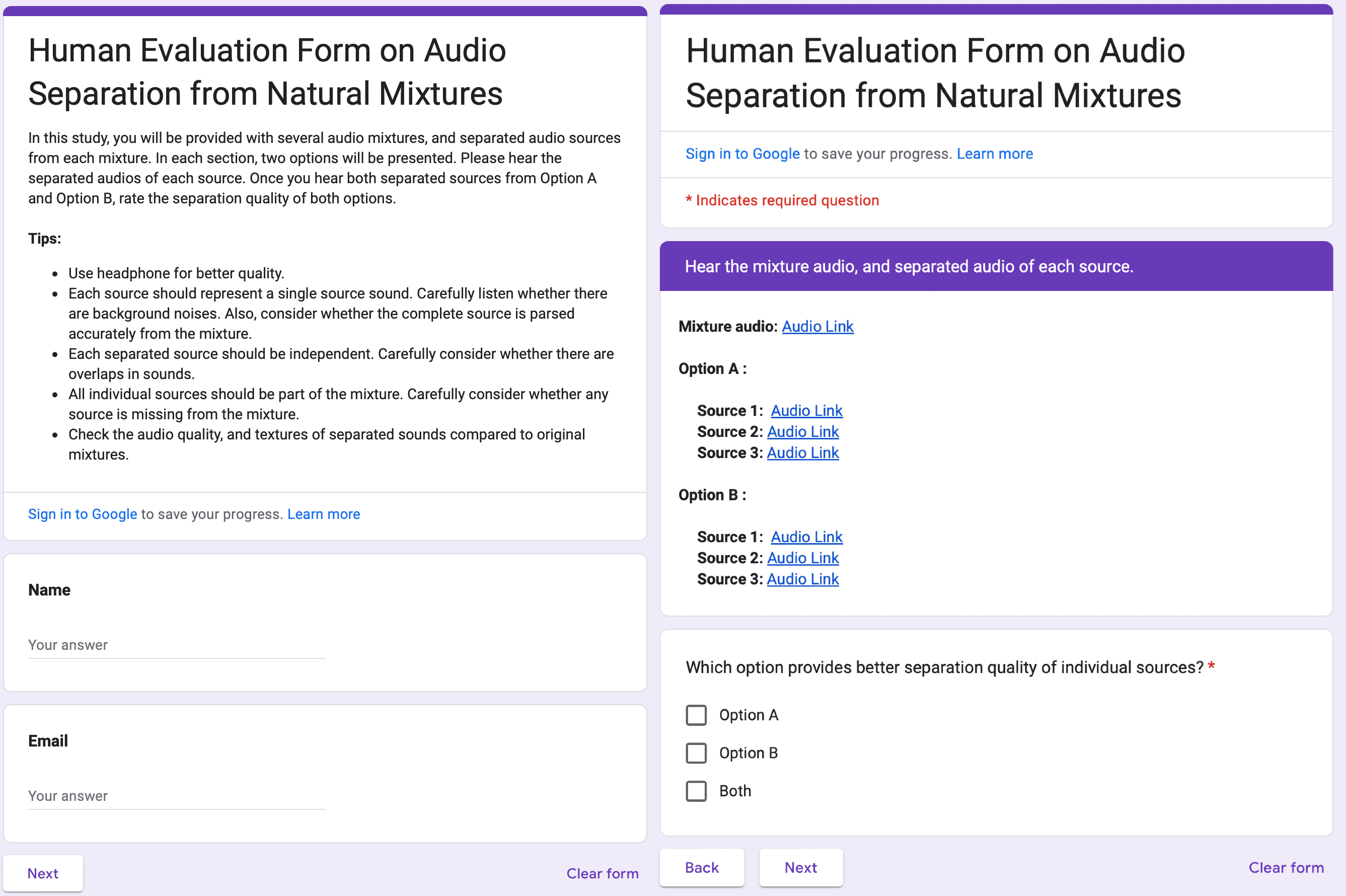}
  \caption{Screenshot of human evaluation form used for the user study. We assess the performance of audio separation from natural mixtures by human evaluation, without having access to any ground truth audios of individual sources. Each user compares the separation quality of two competitive models for several mixtures.}
  \label{fig:7}
  \vspace{10mm}
\end{figure*}

We conduct a user study to analyze the audio separation performance from real-world mixtures, where we don't have access to any ground truth sources. Each user is provided with $20$ mixture samples, and their corresponding separated audios with competitive models. For each sample, user rates the superior separation quality between two choices. We collect the data of wins, loses, and ties across different models using the user study. The results of this study are reported in Table~\ref{tab:user_study_details}. We attach the screenshot of the human evaluation form with detailed guidelines in Fig.~\ref{fig:7}.

\input{tables/source_parsing}
\input{tables/knowledge_parse}

%% file: tables/source_parsing.tex
\begin{table*}[t]
\centering
\caption{Sample of source parsing and generated captions from real-world audio mixtures of AudioCaps dataset. We use ms-CLAP model for audio captioning followed by the source parsing from LLaMA-3-8b with five shot prompts.}
\label{tab:8}
\scalebox{0.82}{
\begin{tabular}{lll}
\toprule
\multicolumn{1}{c}{\textbf{Ground Truth Captions}}                                                                            & \multicolumn{1}{c}{\textbf{Generated Captions}}                                                                                          & \multicolumn{1}{c}{\textbf{Parsed Sources}}                                                                      \\
\midrule
\multirow{2}{*}{\begin{tabular}[c]{@{}l@{}}People talk while an audience laughs \\ and water runs from a faucet\end{tabular}} & \multirow{2}{*}{\begin{tabular}[c]{@{}l@{}}A man and woman are talking and laughing \\ while water runs in the background.\end{tabular}} & \multirow{2}{*}{\begin{tabular}[c]{@{}l@{}}A man talks. A woman talks.\\ People laugh. Water runs.\end{tabular}} \\
                                                                                                                              &                                                                                                                                          &     \\
\midrule
\multirow{2}{*}{\begin{tabular}[c]{@{}l@{}}A child screams, an adult male is \\ talking and vehicles are revving.\end{tabular}} & \multirow{2}{*}{A man speaks while a vehicle engine revs.}                                                                               & \multirow{2}{*}{\begin{tabular}[c]{@{}l@{}}A man speaks.\\ A vehicle engine revs.\end{tabular}}                  \\
                                                                                                                                &                                                                                                                                          &                                                                                                              \\
\midrule
\multirow{2}{*}{\begin{tabular}[c]{@{}l@{}}A child cries and yells \\ and an adult speaks in response\end{tabular}}            & \multirow{2}{*}{A child cries and a woman speaks}                                                                                        & \multirow{2}{*}{\begin{tabular}[c]{@{}l@{}}A child cries.\\ A woman speaks.\end{tabular}}                        \\
                                                                                                                               &  \\ 
\midrule
\multirow{2}{*}{\begin{tabular}[c]{@{}l@{}}A whistle and another man speaks \\ and the whistling stops.\end{tabular}}          & \multirow{2}{*}{A man speaks followed by whistling.}                                                                                     & \multirow{2}{*}{\begin{tabular}[c]{@{}l@{}}A man speaks.\\ A whistling sound.\end{tabular}}                      \\
                                                                                                                               &                                                                                                                                          &                                                                                                                 
\\
\midrule
\multirow{2}{*}{\begin{tabular}[c]{@{}l@{}}A crowd is cheering and shouting, \\ thumping occurs, an adult female speaks.\end{tabular}} & \multirow{2}{*}{\begin{tabular}[c]{@{}l@{}}A crowd is cheering and a woman \\ is speaking.\end{tabular}}                                 & \multirow{2}{*}{\begin{tabular}[c]{@{}l@{}}A crowd is cheering.\\ A woman is speaking.\end{tabular}}             \\
                                                                                                                                       &                                                                                                                                          &                                                                                                                 
\\
\midrule
\multirow{2}{*}{\begin{tabular}[c]{@{}l@{}}A large engine roars and someone \\ is shouts a loud short rout.\end{tabular}}      & \multirow{2}{*}{\begin{tabular}[c]{@{}l@{}}A motor is running while people are \\ shouting.\end{tabular}}                                & \multirow{2}{*}{\begin{tabular}[c]{@{}l@{}}A motor is running.\\ People are shouting.\end{tabular}}              \\
                                                                                                                               &                                                                                                                                          &                                                                                                                 
\\
\midrule
A cat is meowing over and over                                                                                                 & A cat meows and a cat meows.                                                                                                             & A cat meows.                                                                                                    
\\
\midrule
\multirow{2}{*}{\begin{tabular}[c]{@{}l@{}}People are talking and bees are \\ quietly buzzing in the background.\end{tabular}} & \multirow{2}{*}{A bee buzzes and a man speaks.}                                                                                          & \multirow{2}{*}{\begin{tabular}[c]{@{}l@{}}A bee buzzes.\\ A man speaks.\end{tabular}}                            \\
                                                                                                                               &                                                                                                                                          &                                                                                                                 
\\
\midrule
\multirow{2}{*}{\begin{tabular}[c]{@{}l@{}}International music plays as water pours \\ into a pot and finally some splashes\end{tabular}} & \multirow{2}{*}{\begin{tabular}[c]{@{}l@{}}A stream of water is running while \\ music plays in the background.\end{tabular}}            & \multirow{2}{*}{\begin{tabular}[c]{@{}l@{}}A stream of water is running.\\ Music plays in the background.\end{tabular}} \\
                                                                                                                                          &                                                                                                                                          &                                                                                                                        
\\
\midrule
\multirow{3}{*}{\begin{tabular}[c]{@{}l@{}}Several small shuffles of hard objects \\ knocking, followed by loud explosions \\ interspersed by a grown man chuckling.\end{tabular}} & \multirow{3}{*}{A loud bang followed by laughter.}                                                                                       & \multirow{3}{*}{\begin{tabular}[c]{@{}l@{}}A loud bang.\\ Someone laughs.\end{tabular}}                          \\
                                                                                                                                                                                  &                                                                                                                                          &                                                                                                                  \\
                                                                                                                                                                                  &                                                                                                                                          &                                                                                                                 
\\
\midrule
\multirow{2}{*}{\begin{tabular}[c]{@{}l@{}}Birds are flapping their wings\\ and some bird call\end{tabular}}                                                                      & \multirow{2}{*}{\begin{tabular}[c]{@{}l@{}}Birds chirping and flapping \\ their wings.\end{tabular}} & \multirow{2}{*}{\begin{tabular}[c]{@{}l@{}}Birds chirping. \\ Birds  flapping their wings.\end{tabular}} \\
                                                                                                                                                                                  &                                                                                                      &                                                                                                          \\
\midrule
\multirow{2}{*}{\begin{tabular}[c]{@{}l@{}}A toilet flushes and then \\ woman speak in the background\end{tabular}}                                                               & \multirow{2}{*}{\begin{tabular}[c]{@{}l@{}}A toilet flushes followed by a \\ woman speaking.\end{tabular}} & \multirow{2}{*}{\begin{tabular}[c]{@{}l@{}}A toilet flushes.\\ A woman is speaking.\end{tabular}} \\
                                                                                                                                                                                  &                                                                                                            &                                                                 
\\
\bottomrule
\end{tabular}}
\end{table*}

%% file: tables/knowledge_parse.tex
\begin{table*}[ht]
\centering
\caption{Samples of knowledge parsing for various audio sources from the LLM. We use instruction-tune LLaMA-3-8b model for extracting the detailed audio properties of individual sources. We use manually curated 5-shot prompts to guide the LLM for focusing on specific audio properties for assisting the source spearation.}
\label{tab:9}
\scalebox{0.73}{
\begin{tabular}{ll}
\toprule
\textbf{Source}                    & \textbf{Parsed Knowledge}                                                                                                                                                                                                                                                                                                                                                                                                                                                                                                                                                                                                                                                                                                           \\
\midrule
\multirow{4}{*}{\begin{tabular}[l]{@{}l@{}}Alarm clock\\ ringing\end{tabular}} & \multirow{4}{*}{\begin{tabular}[l]{@{}l@{}}The sound of an alarm clock ringing is represented by a sharp, piercing, and repetitive tone, typically in the \\ frequency range of 1-4 kHz, with a characteristic "square-wave" or "sawtooth" shape, featuring a sudden \\ onset, a steady amplitude, often accompanied by a slight decay and a gentle reverberation tail, and a relatively \\ high overall energy level due to the sudden and attention-grabbing nature of the sound.\end{tabular}}                                                                                                                                                                                                                                   \\
                                                                               &                                                                                                                                                                                                                                                                                                                                                                                                                                                                                                                                                                                                                                                                                                                                     \\
                                                                               &                                                                                                                                                                                                                                                                                                                                                                                                                                                                                                                                                                                                                                                                                                                                     \\
                                                                               &   \\
    \midrule
\multirow{5}{*}{Baby laughter}     & \multirow{5}{*}{\begin{tabular}[l]{@{}l@{}}The sound of baby laughter is represented by a series of short, high-pitched, and joyful vocalizations, typically\\ in the frequency range of 200-400 Hz, characterized by a rapid sequence of rising and falling frequencies, with \\ a relatively constant amplitude, often accompanied by subtle variations in pitch and timbre, featuring a "giggly"\\ or "chirpy" quality, and a relatively short duration of around 0.5-1.5 seconds, with a gentle, warm, and intimate \\ quality, and a moderate overall energy level.\end{tabular}}                                                                                                                                             \\
                                   &                                                                                                                                                                                                                                                                                                                                                                                                                                                                                                                                                                                                                                                                                                                                     \\
                                   &                                                                                                                                                                                                                                                                                                                                                                                                                                                                                                                                                                                                                                                                                                                                     \\
                                   &                                                                                                                                                                                                                                                                                                                                                                                                                                                                                                                                                                                                                                                                                                                                     \\
                                   &  \\

    \midrule
\multirow{5}{*}{Cat hissing}       & \multirow{5}{*}{\begin{tabular}[l]{@{}l@{}}The sound of a cat hissing is represented by a high-pitched, sharp, and intense vocalization, typically in the \\ frequency range of 2-4 kHz, with a distinctive "s" or "sh" shape in the spectrogram, featuring a rapid series \\ of short, staccato bursts of energy, often accompanied by a subtle tremolo effect, and a relatively short \\ duration of around 0.1-0.3 seconds, with a moderate to high overall energy level, and a characteristic \\ "attack"and "decay"pattern, with a sudden onset and a gradual release.\end{tabular}}                                                                                                                                          \\
                                   &                                                                                                                                                                                                                                                                                                                                                                                                                                                                                                                                                                                                                                                                                                                                     \\
                                   &                                                                                                                                                                                                                                                                                                                                                                                                                                                                                                                                                                                                                                                                                                                                     \\
                                   &                                                                                                                                                                                                                                                                                                                                                                                                                                                                                                                                                                                                                                                                                                                                     \\
                                   &  \\
    \midrule
    \multirow{5}{*}{Duck quacking}     & \multirow{5}{*}{\begin{tabular}[l]{@{}l@{}}The sound of a duck quacking is represented by a series of loud, sharp, and explosive vocalizations, typically \\ in the frequency range of 200-400 Hz, with a characteristic "honking" quality, featuring a sudden onset and \\ decay, and a relatively constant amplitude, often accompanied by subtle variations in pitch and timbre, and \\ a moderate to high overall energy level, with a distinctive spectral shape and a duration of around 0.5-1.5 \\ seconds, and a possible presence of echoes or reverberations in the environment.\end{tabular}}                                                                                                                            \\
                                   &                                                                                                                                                                                                                                                                                                                                                                                                                                                                                                                                                                                                                                                                                                                                     \\
                                   &                                                                                                                                                                                                                                                                                                                                                                                                                                                                                                                                                                                                                                                                                                                                     \\
                                   &                                                                                                                                                                                                                                                                                                                                                                                                                                                                                                                                                                                                                                                                                                                                     \\
                                   &  \\
\midrule
\multirow{5}{*}{Fox barking}       & \multirow{5}{*}{\begin{tabular}[l]{@{}l@{}}The sound of a fox barking is represented by a series of short, sharp, and high-pitched vocalizations, typically \\ in the frequency range of 1-5 kHz, with a characteristic "yipping" or "yelping" quality, featuring a rapid \\ sequence of rising and falling frequencies, and a relatively constant amplitude, often accompanied by subtle \\ variations in pitch and timbre, and a relatively short duration of around 0.2-0.5 seconds, with a moderate overall\\ energy level, and a distinctive spectral shape featuring a prominent peak in the 2-3 kHz range.\end{tabular}}                                                                                                     \\
                                   &                                                                                                                                                                                                                                                                                                                                                                                                                                                                                                                                                                                                                                                                                                                                     \\
                                   &                                                                                                                                                                                                                                                                                                                                                                                                                                                                                                                                                                                                                                                                                                                                     \\
                                   &                                                                                                                                                                                                                                                                                                                                                                                                                                                                                                                                                                                                                                                                                                                                     \\
                                   &  \\
    \midrule
\multirow{6}{*}{Playing accordion} & \multirow{6}{*}{\begin{tabular}[l]{@{}l@{}}The sound of playing an accordion is represented by a rich, dynamic, and complex mixture of sounds, \\ characterized by a wide frequency range of 50-2000 Hz, featuring a prominent low-frequency foundation\\ (around 50-100 Hz) provided by the instrument's bellows and the diatonic reeds, overlaid with a series of\\ bright, piercing, and harmonically-rich tones produced by the instrument's buttons and keys, with a distinctive \\ "oom-pah" or "chord-like" quality, and often accompanied by subtle vibrato and tremolo effects, as well as \\ occasional percussive attacks and releases, all blending to create a lively, folk-inspired, and emotive sound.\end{tabular}} \\
                                   &                                                                                                                                                                                                                                                                                                                                                                                                                                                                                                                                                                                                                                                                                                                                     \\
                                   &                                                                                                                                                                                                                                                                                                                                                                                                                                                                                                                                                                                                                                                                                                                                     \\
                                   &                                                                                                                                                                                                                                                                                                                                                                                                                                                                                                                                                                                                                                                                                                                                     \\
                                   &                                                                                                                                                                                                                                                                                                                                                                                                                                                                                                                                                                                                                                                                                                                                     \\
                                   &                                                                                                                                                                                                                                                                                                                                                                                                                                                                                                                                                                                                                              \\
\midrule
\multirow{5}{*}{Train whistling}   & \multirow{5}{*}{\begin{tabular}[l]{@{}l@{}}The sound of a train whistling is represented by a distinctive, high-pitched, and piercing tone, typically in \\ the frequency range of 2,000-4,000 Hz, with a sharp attack and decay, and a characteristic "siren-like" shape, \\ often accompanied by a subtle tremolo effect, and a relatively long duration of around 1-3 seconds, with a\\ moderate to high overall energy level, and a sense of spatiality and distance, as if the sound is coming\\ from a specific location.\end{tabular}}                                                                                                                                                                                       \\
                                   &                                                                                                                                                                                                                                                                                                                                                                                                                                                                                                                                                                                                                                                                                                                                     \\
                                   &                                                                                                                                                                                                                                                                                                                                                                                                                                                                                                                                                                                                                                                                                                                                     \\
                                   &                                                                                                                                                                                                                                                                                                                                                                                                                                                                                                                                                                                                                                                                                                                                     \\
                                   &              \\
    \midrule
\multirow{6}{*}{\begin{tabular}[l]{@{}l@{}}Vacuum cleaner \\ cleaning floors\end{tabular}} & \multirow{6}{*}{\begin{tabular}[l]{@{}l@{}}The sound of a vacuum cleaner cleaning floors is represented by a dominant low-frequency hum, typically \\ in the range of 50-200 Hz, with a strong amplitude modulation caused by the motor's rotation and the \\ movement of the vacuum head, often accompanied by a series of high-frequency clicks and rattles from\\ the brushes and wheels, and a gentle whooshing sound from the airflow, with a moderate overall energy level\\ and a relatively consistent spectral shape, punctuated by brief, high-amplitude transients when the vacuum \\ head encounters obstacles or changes direction.\end{tabular}}                                                                      \\
                                                                                           &                                                                                                                                                                                                                                                                                                                                                                                                                                                                                                                                                                                                                                                                                                                                     \\
                                                                                           &                                                                                                                                                                                                                                                                                                                                                                                                                                                                                                                                                                                                                                                                                                                                     \\
                                                                                           &                                                                                                                                                                                                                                                                                                                                                                                                                                                                                                                                                                                                                                                                                                                                     \\
                                                                                           &                                                                                                                                                                                                                                                                                                                                                                                                                                                                                                                                                                                                                                                                                                                                     \\
                                                                                           &
                                                                                           \\
    \midrule
\multirow{6}{*}{Waterfall burbling} & \multirow{6}{*}{\begin{tabular}[l]{@{}l@{}}The sound of a waterfall burbling is represented by a continuous, gentle, and soothing audio signal, featuring \\ a prominent low-frequency energy peak in the range of 20-50 Hz, accompanied by a series of soft, repetitive, \\ and varying pitched tones, typically in the frequency range of 100-500 Hz, with a gradual spectral roll-off \\ towards higher frequencies, and a characteristic "chirping" or "bubbling" quality, often punctuated by \\ occasional, brief, and low-amplitude bursts of energy, and a relatively long duration of several seconds \\ to minutes, with a moderate to high overall energy level.\end{tabular}}                                           \\
                                    &                                                                                                                                                                                                                                                                                                                                                                                                                                                                                                                                                                                                                                                                                                                                     \\
                                    &                                                                                                                                                                                                                                                                                                                                                                                                                                                                                                                                                                                                                                                                                                                                     \\
                                    &                                                                                                                                                                                                                                                                                                                                                                                                                                                                                                                                                                                                                                                                                                                                     \\
                                    &                                                                                                                                                                                                                                                                                                                                                                                                                                                                                                                                                                                                                                                                                                                                     \\
                                    &   \\
\midrule

    \multirow{6}{*}{\begin{tabular}[l]{@{}l@{}}Writing on blackboard \\ with chalk\end{tabular}} & \multirow{6}{*}{\begin{tabular}[l]{@{}l@{}}The sound of writing on a blackboard with chalk is represented by a series of sharp, scratchy, and percussive \\ sounds, typically in the frequency range of 100-500 Hz, with a characteristic "scratch-and-scrabble" texture, \\ featuring a mix of high-amplitude, short-duration events (corresponding to the chalk striking the board) and \\ lower-amplitude, longer-duration events (corresponding to the chalk gliding across the board), often \\ accompanied by subtle variations in tone and timbre depending on the chalk's velocity and angle of incidence,\\  and a relatively low overall energy level due to the soft and dry nature of the writing surface.\end{tabular}} \\
                                                                                             &                                                                                                                                                                                                                                                                                                                                                                                                                                                                                                                                                                                                                                                                                                                                       \\
                                                                                             &                                                                                                                                                                                                                                                                                                                                                                                                                                                                                                                                                                                                                                                                                                                                       \\
                                                                                             &                                                                                                                                                                                                                                                                                                                                                                                                                                                                                                                                                                                                                                                                                                                                       \\
                                                                                             &                                                                                                                                                                                                                                                                                                                                                                                                                                                                                                                                                                                                                                                                                                                                       \\
                                                                                             &   \\
    \bottomrule
\end{tabular}}
\end{table*}